\documentclass[manuscript]{aastex}
\usepackage[T1]{fontenc}

\newcommand{\emails}{lixue@xmu.edu.cn, lujf@xmu.edu.cn}

\shorttitle{Thermally unstable accretion disks. II.}
\shortauthors{Xue et al.}

\begin{document}
\title{Studies of Thermally Unstable Accretion Disks around Black
Holes with Adaptive Pseudospectral Domain Decomposition Method.\\
II. Limit-Cycle Behavior in accretion disks around Kerr black
holes}

\author{Li Xue\altaffilmark{*},
Aleksander S\k{a}dowski\altaffilmark{\dag}, Marek A.
Abramowicz\altaffilmark{\ddag, \dag} and Ju-Fu
Lu\altaffilmark{*}}

\altaffiltext{*}{Department of Physics and Institute of
Theoretical Physics and Astrophysics, Xiamen University, Xiamen,
Fujian 361005, China; \emails}
\altaffiltext{\dag}{Nicolaus
Copernicus Astronomical Center, Bartycka 18, 00-716 Warszawa,
Poland}
\altaffiltext{\ddag}{Department of Physics,
G\"{o}teborg University, SE-412-96 G\"{o}teborg, Sweden}

\begin{abstract}
 For the first time ever, we derive equations governing the
time-evolution of fully relativistic slim accretion disks in the
Kerr metric, and numerically construct their detailed non-stationary
models. We discuss applications of these general results to a
possible limit-cycle behavior of thermally unstable disks. Our
equations and numerical method are applicable in a wide class of
possible viscosity prescriptions, but in this paper we use a
diffusive form of the ``standard alpha prescription'' that assumes
the viscous torque is proportional to the total pressure. In this
particular case, we find that the parameters which dominate the
limit-cycle properties are the mass-supply rate and the value of the
alpha-viscosity parameter. Although the duration of the cycle (or
the outburst) does not exhibit any clear dependence on the black
hole spin, the maximal outburst luminosity (in the Eddington units)
is positively correlated with the spin value. We suggest a simple
method for a rough estimate of the black hole spin based on the
maximal luminosity and the ratio of outburst to cycle durations. We
also discuss a temperature-luminosity relation for the Kerr black
hole accretion discs limit-cycle. Based on these results we discuss
the limit-cycle behavior observed in microquasar GRS 1915+105. We
also extend this study to several non-standard viscosity
prescriptions, including a ``delayed heating'' prescription recently
stimulated by the recent MHD simulations of accretion disks.
\end{abstract}
%---------------------------------------------------------------
%---------------------------------------------------------------
%
\section{Introduction}
%
%---------------------------------------------------------------
This work is motivated by the fundamental issue of how to find the
proper viscosity prescription in the black hole (BH) accretion disk theory.
The ``standard alpha-prescription'', introduced in the seminal
paper by \citet{SS73}, assumes that the viscous
torque ${\mathbb T}$ is proportional to the total pressure $P$,
%---------------------------------------------------------------
\begin{equation}
{\mathbb T} = -\alpha P,
\label{standard-alpha-prescription}
\end{equation}
%---------------------------------------------------------------
where the dimensionless ``viscosity coefficient'' $0 \le \alpha
\le 1$ is a universal phenomenological constant. Hydrodynamical
accretion disk models constructed with the help of
the
%standard prescription (\ref{standard-alpha-prescription}) are
above formula are remarkably successful in describing observed
properties of real disks, but only when these disks are {\it
stationary} and {\it not very luminous}, i.e. when
%---------------------------------------------------------------
\begin{equation}
\frac{\partial}{\partial t} = 0~~{\rm and}~~L < 0.3 \,L_{\rm Edd}.
\label{range-applicability}
\end{equation}
%---------------------------------------------------------------
This is not a surprise. An extension of the Shakura-Sunyaev theory
in terms of the ``slim disk'' models \citep{Abramowicz88}, {\it
predicted} that the properties of accretion disks weakly depend on
$\alpha$ {\it only} in the range (\ref{range-applicability}).
Outside this range, the predicted observables strongly depend on
the form of the viscosity prescription and on the  assumed value
of $\alpha$: the ``small-alpha disks'', $\alpha \approx 0.01$, are
remarkably different from the ``moderate-alpha disks'', $\alpha
\approx 0.1$.
%In the range (\ref{range-applicability})
For luminosities higher than $0.3 \,L_{\rm Edd}$
the
dependence on $\alpha$ is convoluted with the details of radiative
transfer, vertical structure, and general relativistic effects
inside and outside the disk. In addition, the relativistic effects
strongly depend on the BH spin.

These severe technical complications sometimes mask a fundamental problem.
On one hand, with the
help of hydrodynamical models we are able to accurately calculate
observable properties of accretion disks, but only if we assume
({\ref{standard-alpha-prescription}) or other ad hoc formula for
the stress. However, outside the range of applicability (\ref{range-applicability}),
this aproach is questionable and most probably wrong. On the other hand,
the MHD simulations which calculate ${\mathbb T}$ from the first
principles, are not sufficiently realistic today in treating
radiative transfer and other relevant physics accurately enough to
calculate realistic observable properties of accretion disks.

One possible way to make progress is to calculate ``the
state-of-art'' hydrodynamical models with radiative transfer,
vertical structure, general relativistic effects, etc. treated as
accurately as possible and with a properly chosen ``MHD viscosity
prescription'' motivated directly by the MHD simulations. Some
phenomenological parameters that may appear in the MHD prescription,
should be then fixed by a detailed comparison with observations.
This principle was recently used by \cite{Sadowski09} and
\cite{Sadowski10} who calculated a network of models of {\it
stationary} slim accretion disks, covering a wide range of the
relevant parameter space. Our present work depends on these models
in several important aspects. In particular, it takes the stationary
models calculated by S{\k{a}}dowski as the initial condition for the
non-stationary calculations.

We also borrow heavily from the previous paper of our long-term
research program \citep[][hereafter Paper I]{paper_I}. We have
introduced there the \emph{adaptive pseudospectral domain
decomposition} method, which we also use in the present paper.
Results of \cite{paper_I} have been in a good agreement with the
studies on the limit-cycle behavior in the radiation-pressure
supported slim disks by \citet[][hereafter SM01]{SM01}. These, and
all other previous numerical studies of the subject \citep{Honma91,
Lasota-Pelat, SM97, SM98,SM01, Teresi04a,Teresi04b, Mayer06}, have
been performed within the Newtonian hydrodynamics, in which the
Newtonian gravitational potential is replaced by the Paczy{\'n}ski
pseudo-Newtonian potential introduced by \citet{PW80}. It reasonably
models gravity of non-rotating black holes, but fails in the case of
rapidly rotating ones.

Our present paper extends these previous studies by making three new
developments: (i) We perform the first ever fully relativistic {\it
time-dependent} numerical study for a limit-cycle in black hole
accretion disks. (ii) We provide a theoretical tool for
understanding some observational aspects of GRS 1915+105 (and
possibly similar sources) which are unique in showing {\it both} the
evidence for a near-extreme black hole spin, {\it and} the
limit-cycle behavior. (iii) We investigate the possibility of using
the observed properties of the limit cycle to estimate the black
hole spin. These issues are relevant for studying the fundamental
issue of the viscosity prescription.

\section{Basic equations}
In this paper, we consider the axisymmetric relativistic accretion
flows around Kerr BHs. We use the Boyer-Lindquist spherical
coordinates $t$, $r$, $\theta$, $\phi$ to describe the space-time
around a BH. Putting all of the complicated derivations into
Appendix \ref{Appendix_A}, the basic equations which govern the
dynamical behavior of accretion flows can be written as
following\footnote{Throughout this paper we use the $c=G=1$ units.}.
\begin{itemize}

\item Mass conservation:
\begin{equation}
\frac{\partial\Sigma}{\partial t} = -\frac{r\Delta^{1/2}}{\gamma A^{1/2}}
\left[\Sigma\frac{\partial u^t}{\partial t}+\frac{1}{r}\frac{\partial}{\partial r}\left(r\Sigma\frac{V}{\sqrt{1-V^2}}\frac{\Delta^{1/2}}{r}\right)\right],\label{continuity}
\end{equation}
where $\Sigma$ ($=2H\rho$) is the surface density with $H$ being the
half thickness and $\rho$ being the mass density, $V$ is the radial
velocity measured in the corotating frame (CRF), $\gamma$ is the
Lorentz factor, $\Delta=r^2-2Mr+a^2$ and $A= r^4+r^2a^2+2Mra^2$ with
$M$ and $a$ being the BH mass and spin per unit mass, respectively.
The derivative of the contravariant $t$-component of the
four-velocity $\frac{\partial u^t}{\partial t}$ is defined in
equation (\ref{App_eq_dutdt}) as a function of $\frac{\partial
V}{\partial t}$ and $\frac{\partial \cal L}{\partial t}$ (see the
following equations (\ref{radial}) and (\ref{angular}) for the
definitions of these two derivatives).

\item Radial momentum conservation:
\begin{equation}
\frac{\partial V}{\partial t} = \frac{\sqrt{1-V^2}\Delta}{\gamma A^{1/2}}\left[-\frac{V}{1-V^2}\frac{\partial V}{\partial r}+\frac{{\cal A}}{r}-\frac{1-V^2}{\rho}\frac{\partial p}{\partial r}\right], \label{radial}
\end{equation}
where $\cal A$ is defined in equation (\ref{App_def_calA}), $p$
($=\frac{k_{\rm B}}{\mu m_{\rm p}}\rho T+p_{\rm{rad}}$) is the total
pressure consisting of the gas and radiation pressure.

\item Angular momentum conservation:
\begin{equation}
\frac{\partial\cal L}{\partial t} = -\frac{V\Delta}{\gamma\sqrt{1-V^2}A^{1/2}}\frac{\partial\cal L}{\partial r}+
\frac{\Delta^{1/2}}{\gamma\Sigma A^{1/2}}\frac{\partial}{\partial r}\left(\frac{\nu\Sigma A^{3/2}\Delta^{1/2}\gamma^3}{r^4}
\frac{\partial\Omega}{\partial r}\right),\label{angular}
\end{equation}
where $\cal L$(=$u_\phi$) is the angular momentum per unit mass,
$\Omega$ is the angular velocity with respect to the stationary
observer (see equation (\ref{App_def_Omega}) for its definition),
and $\nu$ is the kinematic viscosity coefficient. We evaluate $\nu$
with a diffusive form of the ``standard alpha prescription''
(equation (\ref{standard-alpha-prescription})),
\begin{equation}
\nu=\frac{2}{3}\alpha H \sqrt{\frac{p}{\rho}}. \label{nu}
\end{equation}

\item Half thickness evolution:
\begin{equation}
\frac{\partial H}{\partial t} = -U\cos\Theta_H -\frac{1}{\gamma}\frac{V}{\sqrt{1-V^2}}\frac{\partial H}{\partial r},\label{thickness}
\end{equation}
where $H$ is the half thickness of the disk, $U$ is the vertical velocity, and $\Theta_H$ ($=\arccos\frac{H}{r}$) is the colatitude angle corresponding to the disk surface.

\item Surface vertical motion:
\begin{equation}
\frac{\partial U}{\partial t} = \frac{\Delta^{1/2}}{\gamma^2 A^{1/2}\cos\Theta_H}{\cal R}-
\frac{U}{\gamma^2}\left(\frac{V}{(1-V^2)^2}\frac{\partial V}{\partial t}+
\frac{{\cal L}r^2}{A}\frac{\partial\cal L}{\partial t}\right)-
\frac{U}{H}\frac{\partial H}{\partial t}, \label{vertical}
\end{equation}
where $\cal R$ is defined in equation (\ref{App_def_calR}).
Equations (\ref{thickness}) and (\ref{vertical}) describe the
evolution of the disk surface. This kind of vertical treatment was
first introduced by \citet{SM01} in order to mimic some advantageous
features of a two-dimensional model in a one-dimensional
\citep[see][]{PS94} study. Thus, our study also exhibits some
two-dimensional features (e.g., the vertical component of the
velocity at the disk surface) so that we prefer to call it
1.5-dimensional.

\item Energy conservation:
\begin{equation}
\frac{\partial T}{\partial t} = \frac{1}{\Sigma}\frac{r\Delta^{1/2}}{\gamma A^{1/2}}\left[
\frac{F^+-F^-}{c_V}+(\Gamma_3-1)T\Sigma\left(-\frac{\partial u^t}{\partial t}-\frac{1}{r^2}\frac\partial{\partial r}(r^2u^r)\right)\right]-\frac{V\Delta}{\gamma\sqrt{1-V^2}A^{1/2}}\frac{\partial T}{\partial r}, \label{energy}
\end{equation}
where $T$ is the temperature of gas, $F^+$ is the local viscous heat
generation rate (equation (\ref{App_def_F^+})), and $F^-$ is the
radiative cooling rate given by the bridge formulae, which is valid
for both optically thick and thin regimes,
\begin{equation}\label{bridging_formulae}
F^-=\frac{8\sigma T^4}{3\tau_{\rm{R}}/2+\sqrt{3}+1/\tau_{\rm{P}}},
\end{equation}
where $\tau_{\rm{R}}$ and $\tau_{\rm{P}}$ are the Rosseland and Planck mean optical depths \citep[see e.g.,][]{Abramowicz96}.
%(see the equations (\ref{F^-}) and (\ref{Aeq:Fm})).
The corresponding radiation pressure is given by,
\begin{equation}
p_{\rm{rad}}=\frac{F^-}{2}\left(\tau_{\rm{R}}+\frac{2}{\sqrt{3}}\right).
\end{equation}
$c_{\rm V}$ and $\Gamma_3$ are defined in equations
(\ref{App_def_cV}) and (\ref{App_def_Gamma3}), respectively.
\end{itemize}

%These basic equations (\ref{continuity}), (\ref{radial}), (\ref{angular}), (\ref{thickness}), (\ref{vertical}), and (\ref{energy}) are the same as the equations
%(\ref{App_eq_dSigmadt}), (\ref{App_eq_dVdt}), (\ref{App_eq_dLdt}), (\ref{App_eq_dHdt}), (\ref{App_eq_dUdt}), and (\ref{App_eq_dTdt}) respectively. We solve them
%for the six essential variables $\Sigma$, $V$, $\cal L$, $H$, $U$, and $T$, which describe the structure of disks.

\noindent In this paper, the properties of an accretion disk depend on three crucial parameters:
\begin{itemize}
\item Dimensionless black hole spin $a^*$:
\begin{equation}
a^*\equiv\frac{a}{GM/c}=\frac{a}{M},~~~-1\leq a^* < 1;
\end{equation}
\item Diffusive viscosity parameter $\alpha$ (see equation (\ref{nu})):\\
 $$0<\alpha<1;$$
\item Dimensionless mass-supply rate $\dot m$:\\
\begin{equation}
\dot m\equiv\frac{\dot M(r_{\rm{out}})}{\dot M_{\rm{Edd}}}.
\end{equation}
\end{itemize}
Where $\dot M(r_{\rm{out}})$ and $\dot M_{\rm{Edd}}\equiv 64\pi GM/(c \kappa_{\rm{es}})$ are the accretion rate at the outer boundary $r=r_{\rm{out}}$ and the Eddington accretion rate (corresponding to the Eddington luminosity for a non-rotating BH), respectively.

\section{Numerical method}
Following Paper I, we use \emph{the adaptive pseudospectral domain
decomposition method} to solve the basic equations. This method has
already been validated by the successful reproduction of the
limit-cycle behavior in the pseudo-Newtonian framework. The
relativistic time-dependent code used in this work is an upgraded
version of the previous pseudo-Newtonian one. For the initial
conditions, we interpolate the relativistic stationary global slim
disk solution of \citet{Sadowski09} on the grid of the
time-dependent code. We study the accretion disk around a BH with
mass $M=10M_{\odot}$ and setup the computation domain from a radius
between the BH horizon and ISCO (Innermost Stable Circular Orbit),
say $r=2.5r_{\rm{g}}$ for a nonrotating BH
($r_{\rm{ISCO}}=3r_{\rm{g}}$), to $10^3r_{\rm{g}}$
($r_{\rm{g}}=2GM/c^2=2M$). As in Paper I, the computational domain
includes a highly supersonic region, with high radial velocities
(about $0.2\sim0.3c$ at the inner boundary). At this inner
boundary, we adopt the free-type boundary conditions, i.e., we let
all variables evolve naturally according to their basic equations
except $\cal L$, which is set by an extrapolation of the nearest two
points. This kind of inner boundary conditions, in practice, are
quite stable because the accretion flow always supersonically
inflows beyond the inner boundary without numerical spurious
reflection. Meanwhile, they are also more natural and physical than
directly setting the viscous torque to vanish outside the BH's
horizon. At the outer boundary, we fix $V$ and $\Sigma$ to ensure
constant mass supply, and also fix $\cal L$ and set $U=0$ to avoid
the numerical spurious evolution.

We have performed computations for two particular cases to check the
relativistic code. Both consist of an accretion disk with
$\alpha=0.07$ and $\dot m=0.02$, while the BH spin was set to
$a^*=0$ and $a^*=0.95$, respectively. The essential difference
between these two cases is in the thermal stability. According the
local analysis of stability \citep[e.g.,][p.155]{Kato_book}, only
the non-spinning case is thermally stable (see Fig. \ref{fig:beta}).
In Figure \ref{fig:DensTemp}, we show the profiles of surface
density and temperature for both cases. It is clear that for the
$a^*=0$ case (left panel), the disk remains in the steady state for
more than $1.2\times10^4$ seconds, therefore it can be regarded as a
stable disk (the tiny differences between the initial state and the
snapshot are due to the differences of numerical methods used for
obtaining stationary and time-dependent solutions and are considered
not significant). On the contrary, for the highly spinning case
(right panel) the disk undergoes an outburst immediately after the
beginning of simulation. These results validate the ability of the
code to distinguish the thermally stable and unstable behavior, and
also prove that the BH spin significantly influences the thermal
stability of the accretion disk, which motivates us to explore the
limit-cycle behavior of thermally unstable disks around Kerr BHs.

\section{Numerical exploration}\label{Sec_NumExplo}
\subsection{Parameter space and characteristic quantities}
In order to understand the effects of BH spin on the limit-cycle
behavior, it is necessary to investigate the parameter space
($a^*,\alpha,\dot m$) with a series of models. In this section, we
describe and discuss results of twelve runs assuming different sets
of the input parameters span on the grid: $a^*=(0, 0.5, 0.95)$,
$\alpha=(0.07, 0.1)$, and $\dot{m}=(0.06, 0.1)$ (cf. Table
\ref{tab1}). All of the cases are thermally unstable and undergo
recursive limit-cycle evolution.

To describe the limit-cycle behavior, we define four characteristic
quantities, which are the Outburst Duration (the full width at half
maximum of light-curve, hereafter OD), Cycle Duration (time interval
between two outbursts, hereafter CD), ratio of OD to CD, and Maximal
Luminosity (the peak value of the light curve during outburst,
hereafter ML; calculated intrinsically, with no ray-tracing),
respectively. In Figure \ref{fig:sketch}, we visualize these
definitions on a sketched light-curve.

The computations for each one of those twelve cases continue until
the CD converges to a constant value. The constancy of CDs implies
that the computations has reached a state uniquely determined by the
parameters ($a^*,\alpha,\dot m$), without any impact of the initial
conditions. In practice, the required level of convergence is
achieved after three or five cycles. We have performed computations
for each case for more than seven cycles (some cases reached ten
cycles). Our analysis below is based on the last four cycles only.

In Table \ref{tab1} we list the mean values, standard and relative
deviations of each characteristic quantity for our twelve models.
These values are calculated based on the retained cycles only. Since
the relative deviations (the percentages in the brackets) are all
very low, we do not show the error bars for the data points in the
subsequent figures.

\subsection{Impact of BH spin}
In order to reveal the effects of BH spin on limit-cycle behavior,
we have divided our twelve models into four groups to reflect the
different effects of viscosity and mass supply (i.e., mass accretion
rate at the outer boundary). In Figures \ref{fig:4plots} and
\ref{fig:ML_ODC}, the models with ($\alpha=0.07, \dot m=0.06$) and
different spins are represented by empty stars, while those with
($\alpha=0.1, \dot m=0.06$), ($\alpha=0.07, \dot m=0.1$), and
($\alpha=0.1, \dot m=0.1$) by filled stars, empty circles, and
filled circles, respectively (both are based on the data from Table
\ref{tab1}).

In the four panels of Figure \ref{fig:4plots} we plot OD, CD, OD/CD
and ML as a function of $a^*$. It is clear that there is no distinct
and consistent dependence on $a^*$ neither for OD nor CD (panels (a)
and (b)). The ratio of OD to CD (panel (c)) is almost independent of
$a^*$ for all sets of input parameters. On the contrary, ML exhibits
a perfect correlation with $a^*$ (panel (d)), though this relation
is still slightly distorted by the impact of $\alpha$ and $\dot m$.

These facts may be easily understood in the framework of the general
relativity. The effects of BH spin are restricted to the innermost
region of accretion disks. Therefore, the quantity ML, which
corresponds mostly to the radiation emerging from the inner part of
an accretion disk, must display a strong dependence on $a^*$. The
other quantities (OD, CD and OD/CD) don't have such a strong
dependence on $a^*$ as they depend on the disk structure and its
evolution at a wide range of radii (from the inner part up to more
than $100r_{\rm{g}}$), where the effects of viscosity and
mass-supply dominates over the effect of BH spin.

\subsection{Possibility of estimating the black hole spin with limit-cycle}
As has been discussed in the previous section, the effects
of BH spin on the quantities OD and CD are easily disturbed by the
complex effects of viscosity and mass-supply; while the ratio OD/CD
has no discernible dependence on BH spin. Thus, they cannot serve
as proper probers of BH spin. Even ML, which has monotonic and
positive correlation with $a^*$, cannot be used directly to estimate
the BH spin, because the dependence of ML on BH spin is
significantly affected by other factors (see panel (d) of Figure
\ref{fig:4plots}).

None of those quantities can be directly used as a probe,
but there still exists the possibility of probing the BH spin with
them. In fact, we find that the combination of ML and OD/CD can be
used to estimate the BH spin. In Figure \ref{fig:ML_ODC}, we show
the OD/CD-ML diagram for all the models. The ratio OD/CD raises with
increasing values of $\alpha$ and $\dot m$, and is hardly dependent
on $a^*$ (see panel (c) of Figure \ref{fig:4plots}). It is
remarkable that the combined effects of viscosity and mass-supply
make ML scale almost linearly with the ratio OD/CD. For a given BH
spin there is a single straight line reflecting the dependence of ML
on OD/CD. These lines do not intercept and have similar slopes.
Therefore, the OD/CD-ML diagram may be used for a rough estimation
of BH spin if only the limit cycle parameters can be obtained from
observational data.

We find that the dependence of ML on OD/CD may be approximated by the following formula
\begin{equation}
\label{mlvsodcd_fit}
 {\rm ML}\,/\,L_{\rm{Edd}}=7.59\,\, {\rm OD/CD} + 0.71 + 7.18\, \eta,
\end{equation}
where $\eta$ is the efficiency of accretion for a thin disk and is given by
$$\eta=1-\sqrt{1-\frac{2 M}{3 r_{\rm ms}}},$$
with $r_{\rm ms}$ being the radius of the marginally stable orbit.
In Figure \ref{fig:ML_ODC} we plot with dashed lines the fits
obtained with these formulae for $a^*=0,\,0.5$ and $0.95$. Given the
values of ML and OD/CD, one can easily obtain  the radius of the
marginally stable orbit from Eq.~\ref{mlvsodcd_fit}, and
subsequently the BH spin.

However, there are at least a few limitations for the
application of this method. The first one is related to the fact
that the OD/CD-ML relation is still subject to a significant
dispersion (see Fig. \ref{fig:ML_ODC}). For a given BH spin, the
combined impacts of viscosity and accretion rate result in a
relation which is only roughly linear.

The second one is related to the modulation of the emitted
radiation by the gravitational redshift and focusing effects near
the BH \citep[e.g.][]{CunninghamBardeen73}. In our method, the ratio
OD/CD is measured in the Boyer-Lindquist (observer's) time, but ML
is intrinsic, i.e., it is the integral of local radiation flux in
the whole disk. Thus, one should calculate ML only after calculating
the observed disk spectrum, which is subject to the gravitational
effects mentioned above. It would be possible by applying
ray-tracing techniques \citep[e.g.][]{Fanton97} to solve this issue.
However, the calculations will be very complicated and out of the
scope of our paper. This is because the precise calculation of the
emitting spectrum would require knowledge about the intrinsic local
emission which cannot be approximated as a perfect black body,
especially in effectively optically thin regions. Even so, we can
still anticipate that the apparent MLs would be lower than the
intrinsic ones (even for the face on case) due to the reduction of
the apparent luminosity of disks by the gravitational redshift.
However, this decrease should not be strong, as most of the flux
contributing to ML comes from a region extending up to $\sim10r_g$
where the gravitational reddening has little impact. We argue that
the observed value of ML is not expected to be lower than the
intrinsic one by more than a few percent. The observed values of ML
would be further decreased if the inclination is not face on --- one
should account for this fact before applying the method.

The last one is the model-dependence of our method. We assume the
limited classical theory to describe the accretion flow. We neglect
mass outflows, magnetic fields and apply the $\alpha$ prescription
of viscosity with $\alpha$ independent of radius. These factors,
especially the viscosity treatment, may have significant impact on
ML vs OD/CD relation that we have obtained. \cite{Nayakshin00} show
that reproducing the light curves of GRS 1915+105 is possible only
if much more complicated models are considered. Nevertheless, if the
limit-cycle behavior is the actual reason of the observed
variability of some microquasars, and if the extraction of ML and
OD/CD parameters from the light curves is possible, then, under the
limitations of our model, the BH spin can be roughly estimated,
before other more accurate methods, such as fitting spectral energy
distribution, are applied.

\subsection{Evolution of the temperature-luminosity relation for the limit-cycle in classical theory}
The maximal disk temperature-luminosity ($T$-$L$) correlation
predicted by the classical  theory of limit-cycles has been used to
compare with observations
\citep[e.g.][]{Gierlinski04,Kubota04,Kubota01}. In Figure
\ref{fig:L_MT}, we show the $T$-$L$ evolution for the last cycle
calculated in each model. The panels present the evolution for a
given set of $\alpha$ and $\dot m$ and different BH spins ($a^* =
0,\,0.5$ and $0.95$ marked by solid, dashed and dotted lines,
respectively). These curves reveal several common features: (1) All
of the cycles begin at the closely-located triangle points and spend
a quarter of CD evolving along the curve in the anti-clockwise
direction before they reach the square points, and finally spend the
other $3/4$ of CD to return to the starting location (triangles);
(2) The evolution of all the cycles fits (with exception of the
outbursts of the $a^*=0.95$ case) between two straight lines
corresponding to $L_{\rm disk}\propto T^4_{\rm max}$ relations with
different color correction factors \citep[$f_{\rm col}=1$ and $7.2$,
see][]{Gierlinski04}; (3) The $T$-$L$
 curves of all the cycles are similar for most of CD except for the outburst state --- the BH spin impact is revealed at the outburst of limit-cycle
 only; (4) During the mass restoring process (prior to the next outburst) the disk obeys $L_{\rm disk}\propto T^{0.7}_{\rm max}$  (see the dashed straight
 lines in all panels).

\section{Discussion}

The theory based on the standard viscosity prescription
(\ref{standard-alpha-prescription}) predicts that
%for nearly Eddington
%luminosities, $0.3 \le L/L_{\rm Edd} \le 1.0$, accretion disks
%should be radiation pressure supported, $P = P_{\rm gas} + P_{\rm
%rad} \approx P_{\rm rad}$,
radiation pressure-supported regions of radiatively efficient
accretion disks are thermally and viscously unstable. The range of
luminosity within which the disk is unstable is narrower for an (ad
hoc) alternative viscosity prescription
%---------------------------------------------------------------
\begin{equation}
{\mathbb T} = -\alpha \sqrt {P^{2 -\mu} P^{\mu}_{\rm gas}} ,
\label{alternative-alpha-prescription}
\end{equation}
%---------------------------------------------------------------
where $0 \le \mu \le 1$ is a constant parameter
\citep{Szuszkiewicz90}, with $\mu = 0$ corresponding to the standard
prescription and $\mu = 1$ to the so-called ``geometrical mean''
prescription. The instability disappears for $\mu > 8/7$. A recent
important work by \cite{Hirose09} shows that the standard
Shakura-Sunyaev viscosity prescription
(\ref{standard-alpha-prescription}) fits their MHD simulations of
accretion disks much better than the prescriptions
(\ref{alternative-alpha-prescription}) with $\mu > 0$. However,
\cite{Hirose09} also find that the radiation pressure supported MHD
disks are thermally {\it stable} in their simulations (it is not
known whether they are viscously stable). This result contradicts
the hydrodynamical stability analysis. What are the MHD effects that
stabilize the disk? One explanation is given in \cite{Hirose09}.
They argue that although the viscosity prescription has the form
(\ref{standard-alpha-prescription}), it should nevertheless be
modified in a subtle way, because the ``viscous heating'' ${\mathbb
Q}^+$ that occurs in the disk is {\it delayed} with respect to the
MHD stress ${\mathbb T} = {\mathbb T}_{\rm MHD}$,
%---------------------------------------------------------------
\begin{equation}
{\mathbb Q}^+(t) = {\mathbb T}(t - \Delta t)\left[
\frac{d\Omega}{dr}\right ].
\label{delayed-heating}
\end{equation}
%---------------------------------------------------------------
Here $\Delta t$ is the delay, found to be about 10 dynamical times,
and $\Omega$ is the angular velocity of matter. It should be obvious
that this modification is irrelevant in the stationary case. In an
exhaustive {\it Newtonian} analytical stability analysis
(\cite{Ciesielski11}, see also \cite{Lin2011}), parallel to the
numerical work described in the present paper, we have confirmed (in
general) that in some parameter space the suggestion made in
\cite{Hirose09} does stabilize the radiation pressure thermal
instability. However, we also found a variety of different parameter
ranges with interesting, and complex, oscillatory behaviors that
need to be further investigated. The present paper makes the first
step into this direction, by investigating the standard case $\Delta
t = 0$. We already numerically simulate a few representative cases
with $\Delta t \not = 0$. Another explanation to the absence of the
thermal instability is given by \cite{Zheng2011}. They attribute
the stability to the magnetic pressure in the accretion disk, which
was neglected in the traditional hydrodynamical stability analysis.
They show that if the magnetic pressure decreases with response to
an increase of temperature of the accretion flow, the threshold of
accretion rate above which the disk becomes unstable increases
significantly compared to the case of not considering the magnetic
pressure. The physical reason is that in this case the dependence of
turbulent dissipation heating on temperature becomes weaker.

On the observational front, spin of BHs in several galactic BH
candidates have recently been measured by fitting the observed
spectral energy distribution with the relativistic geometrically
thin accretion disk model \citep{Shafee06,McClintock06}. Among these
objects, GRS 1915+105 is the most special one. It is believed to
have an extreme Kerr BH with the spin very close to $a^*=J/M^2=1$
\citep[see][]{McClintock06}. It is also the only object that
exhibits the quasi-regular luminosity variations \citep{Belloni97,
Nayakshin00, Janiuk02, Watarai03, Ohsuga06, Kawata06,
janiukczerny11}, similar to the limit-cycle predicted by the
standard {\it Newtonian}, hydrodynamical models. Several questions
need to be investigated here. May one explain the luminosity
variations observed in GRS 1915+105 as the classic ``slim-disk''
limit-cycle behavior? What is the role of the nearly extreme spin of
GRS 1915+105 in the context of the observed variability? Could the
case of GRS 1915+105 help to find the answer about the proper
viscosity prescription?

Our research shows that the shape of light curves weakly depend on
BH spin. The only quantity that feels the impact of the BH spin is
the maximal luminosity. Therefore, under the assumptions adopted in
this paper (e.g., a constant $\alpha$), the limit cycle can produce
light curves which are only qualitatively similar to those observed
in GRS1915+105, independent of BH spin. To obtain better agreement
one has to construct more complicated models
\citep[e.g.,][]{Nayakshin00}. We prove that the cycle duration
weakly depends on the BH spin. Thus, the observed value for GRS1915+105
most probably results from combined effects of
viscosity and mass-supply rate and is not affected by the BH spin. The
very presence of limit cycle variability only in this object, if it
is true, may challenge the explanation given by \cite{Hirose09}
, but could be understood in the model of \cite{Zheng2011}
as due to the high luminosity of this source compared to other
sources.

\section{Summary and conclusions}
This paper is a sequel of Paper I \citep[]{paper_I}. We have changed
the framework of study from the previous pseudo-Newtonian
hydrodynamics to relativistic one. We have established the
time-dependent basic equations for slim disk accretion in the full
general relativity (Appendix \ref{Appendix_A}). We continue to use
the \emph{adaptive pseudospectral domain decomposition} method,
which has been validated in Paper I, to solve these new equations.
Our main results and conclusions are summarized as follows.

1. We have calculated two models assuming $\alpha=0.07$ and $\dot
m=0.02$ and two values of BH spin. We find that the model with a
non-spinning BH ($a^*=0$) can stay in the steady state, but the
model with a high-spinning BH ($a^*=0.95$) undergo an outburst
immediately after the beginning of the computation. These features
agree with the predictions of the classical local analysis of
stability (see Figure \ref{fig:beta}), thus the ability of the new
relativistic code for distinguishing the thermally stable and
unstable disks is validated, and it is confirmed that the BH spin
can change the thermal stability of an accretion disk.

2. Following the time-evolution of twelve models we have explored
the effects of BH spin on the limit-cycle behavior. We define four
characteristic quantities to describe the limit-cycle behavior: the
outburst duration (OD), cycle duration (CD), ratio of OD to CD, and
maximal luminosity (ML). We find that the effects of BH spin on the
quantities that depend on the disk structure at a wide range of
radii (OD, CD and OD/CD), are easily disturbed and even obscured by
the combined effects of viscosity and mass-supply. However, ML is an
exception. As it depends on the radiation coming from the inner
region of the disk, it overcomes the impact of viscosity and
mass-supply and reveals a monotonous positive dependence on BH spin.

3. We have discussed the possibility of using the OD/CD-ML diagram
to estimate the BH spin. The advantage of this method is its
simplicity, but the dispersion related to the unknown viscosity and
mass-supply, as well as its model dependence limit its application.
We suggest that one can use the OD/CD-ML relation to perform rapid
and rough estimation, before more accurate BH spin estimations based
on other methods are available.

4. We have presented and discussed the evolution of the $T$-$L$
relation for limit-cycles with different model parameters, proving
that BH spin changes the disk evolution only during the outburst
phase.

\acknowledgments

We thank Feng Yuan for beneficial discussion. This work was
supported by 973 Program under grant 2009CB824800, the National
Natural Science Foundation of China under grants 10833002 and
11003016, the Natural Science Foundation of Fujian Province of China
under grant 2010J01017, and by Polish Ministry of Science grants
N203 0093/1466, N203 304035, N203 380336 and N203 381436.

\appendix

\section{Derivation of basic equations}\label{Appendix_A}
The basic equations used in our paper, which make exploring the
time-dependent behavior of accretion disks around Kerr BHs possible,
are derived basing on \citet{Abramowicz96,Abramowicz97},
\citet{gammiepopham} and \citet{Sadowski09}. In this appendix, we
give all of the derivation details that are ignored in the main part
of our paper.

\subsection{Metric and four-velocity}
We use the Boyer-Lindquist spherical coordinates $t$, $r$, $\theta$,
$\phi$ to describe the Kerr BH metric. We use its reduced form near
the equatorial plane, because the disk equations involve vertically
averaged quantities and the disks we deal with are not very thick.
The nonvanishing covariant and contravariant components of the
reduced metric can be written as,
\begin{equation}
g_{tt}=-\frac{r-2M}{r},\ g_{t\phi}=-\frac{2Ma}{r},\ g_{\phi\phi}=\frac{A}{r^2},\ g_{rr}=\frac{r^2}{\Delta},\ g_{\theta\theta}=r^2;
\end{equation}
and
\begin{equation}
g^{tt}=-\frac{A}{r^2\Delta},\ g^{t\phi}=-\frac{2Ma}{r\Delta},\ g^{\phi\phi}=\frac{r-2M}{r\Delta},\ g^{rr}=\frac{\Delta}{r^2},\ g^{\theta\theta}=\frac{1}{r^2}.
\end{equation}
Where $M$ and $a$ are the mass and specific angular momentum of the BH respectively, $\Delta\equiv r^2-2Mr+a^2$, $A\equiv r^4+r^2a^2+2Mra^2$. The metric is in the geometrical units ($c=G=1$) and its signature is $(-+++)$.

The stress-energy tensor reads,
\begin{equation}
T^{ik}=\rho u^iu^k+pg^{ik}+s^{ik}+u^kq^i+u^iq^k,
\end{equation}
where $\rho$, $u^i$, $p$,  $s^{ik}$ and $q^i$ are the rest mass density, contravariant components of four-velocity, total pressure (the sum of the gas and radiation pressure), viscous stress tensor, and radiative energy flux, respectively.

The general form of the four-velocity is,
\begin{equation}\label{general_4V}
u^i=\gamma\left(e^i_{(t)}+V^{(r)}e^i_{(r)}+V^{(\phi)}e^i_{(\phi)}+V^{(\theta)}e^i_{(\theta)}\right),
\end{equation}
where $\gamma$ is the Lorentz factor, $V^{(j)}$ are velocities as measured in the Local Non Rotating Frame (LNRF) and $e_{(j)}$ are LNRF basis vectors. Near the equatorial plane, they take the following forms,
\begin{equation}\label{def-gamma}
\gamma^{-2}=1-\left(V^{(r)}\right)^2-\left(V^{(\phi)}\right)^2;
\end{equation}
\begin{mathletters}\label{V^i}
\begin{eqnarray}
\label{V^r} V^{(r)} &=& \frac{1}{\gamma}\frac{V}{\sqrt{1-V^2}},\\
\label{V^theta} V^{(\theta)} &=& U\cos{\theta},\\
V^{(\phi)} &=& \frac{1}{\gamma}\frac{{\cal L}r}{A^{1/2}};
\end{eqnarray}
\end{mathletters}
and
\begin{mathletters}\label{e_i}
\begin{eqnarray}
\label{e_t} e_{(t)}&=&\frac{A^{1/2}}{r\Delta^{1/2}}\frac{\partial}{\partial t}+\frac{2Ma}{A^{1/2}\Delta^{1/2}}\frac{\partial}{\partial \phi},\\
e_{(r)}&=&\frac{\Delta^{1/2}}{r}\frac{\partial}{\partial r},\\
e_{(\theta)}&=&\frac{1}{r}\frac{\partial}{\partial \theta},\\
\label{e_phi} e_{(\phi)}&=&\frac{r}{A^{1/2}}\frac{\partial}{\partial \phi},
\end{eqnarray}
\end{mathletters}
where we have followed \citet{BPT72} to use the relations (\ref{e_t})-(\ref{e_phi}), and we have introduced the radial velocity measured in the corotating frame (CRF) $V$,  the vertical velocity parameter $U$ and the angular momentum per unit mass $\cal{L}$($\equiv u_{\phi}$) to describe accretion flows.

According to equations (\ref{V^i}), (\ref{e_i}) and (\ref{general_4V}), the contravariant components of the four-velocity in terms of $V$, $U$ and $\cal{L}$ are
\begin{mathletters}
\begin{eqnarray}
\label{u^t} u^t&=&\gamma\frac{A^{1/2}}{r\Delta^{1/2}},\\
%u^r&=&\gamma V^{(r)}\frac{\Delta^{1/2}}{r}=\frac{V}{\sqrt{1-V^2}}\frac{\Delta^{1/2}}{r},\\
%u^\phi&=&\gamma\frac{2Ma}{A^{1/2}\Delta^{1/2}}+\gamma V^{(\phi)}\frac{r}{A^{1/2}}=\frac{{\cal L}r^2}{A}+\gamma\omega\frac{A^{1/2}}{r\Delta^{1/2}},\\
%u^\theta&=&\gamma V^{(\theta)}\frac{1}{r}=\frac{\gamma}{r}U\cos{\theta}.
\label{u^r} u^r&=&\frac{V}{\sqrt{1-V^2}}\frac{\Delta^{1/2}}{r},\\
u^\phi&=&\frac{{\cal L}r^2}{A}+\gamma\omega\frac{A^{1/2}}{r\Delta^{1/2}},\\
u^\theta&=&\frac{\gamma}{r}U\cos{\theta}.
\end{eqnarray}
\end{mathletters}
The relevant covariant components are
\begin{mathletters}
\begin{eqnarray}
u_t&=&-\gamma\frac{r\Delta^{1/2}}{A^{1/2}}-\omega{\cal L},\\
u_r&=&\frac{r}{\Delta^{1/2}}\frac{V}{\sqrt{1-V^2}},\\
\label{u_phi} u_\phi&=&\cal L,\\
u_\theta&=&\gamma rU\cos{\theta}.
\end{eqnarray}
\end{mathletters}
At last, according to equation (\ref{def-gamma}), the Lorentz factor can be written as,
\begin{equation}
\gamma^{2}=\frac{1}{1-V^2}+\frac{{\cal L}^2r^2}{A}.
\end{equation}

\subsection{Mass conservation}
The general form of the continuity equation is
\begin{equation}
\nabla_i(\rho u^i)=0.
\end{equation}
After vertical integration it takes the following form
\begin{equation}
u^t\frac{\partial\Sigma}{\partial t}=-\frac{1}{r}\frac{\partial}{\partial r}(\Sigma ru^r)-\Sigma\frac{\partial u^t}{\partial t},
\end{equation}
where
$\Sigma$
%($\equiv2r\cos{\Theta_H}$, see below \S\ref{App_section_HalfOpen} for the definition of $\Theta_H$)
is the surface density. Substituting $u^t$ and $u^r$ with relations (\ref{u^t}) and (\ref{u^r}), we have
\begin{equation}\label{App_eq_dSigmadt}
\frac{\partial\Sigma}{\partial t}=-\frac{r\Delta^{1/2}}{\gamma A^{1/2}}
\left[\Sigma\frac{\partial u^t}{\partial t}+\frac{1}{r}\frac{\partial}{\partial r}\left(r\Sigma\frac{V}{\sqrt{1-V^2}}\frac{\Delta^{1/2}}{r}\right)\right],
\end{equation}
where the time derivative $\frac{\partial u^t}{\partial t}$ can be substituted with the equation
\begin{equation}\label{App_eq_dutdt}
\frac{\partial u^t}{\partial t}=\frac{A^{1/2}}{r\Delta^{1/2}}\frac{1}{\gamma}\left[\frac{V}{(1-V^2)^2}\frac{\partial V}{\partial t}+
\frac{{\cal L}r^2}{A}\frac{\partial\cal L}{\partial t}\right],
\end{equation}
which is obtained by operating time derivative on equation (\ref{u^t}). In the right-hand-side of the above equation, there still exist two time derivatives $\frac{\partial V}{\partial t}$ and $\frac{\partial\cal L}{\partial t}$, but they can be evaluated before the evaluation of $\frac{\partial u^t}{\partial t}$ with equations (\ref{App_eq_dVdt}) and (\ref{App_eq_dLdt}).

\subsection{Radial momentum conservation}
The radial momentum conservation is expressed as the vanishing of the $r$-component of the divergence of the stress-energy tensor
\begin{equation}
\nabla_i T^{ir}=0.
\end{equation}
Expanding it
\begin{eqnarray}
\nonumber && u^i\nabla_i u^r+\frac{1}{\rho}g^{rr}\frac{\partial p}{\partial r}=0\\
\Rightarrow && u^t\frac{\partial u^r}{\partial t}+u^r\frac{\partial u^r}{\partial r}+u^i\Gamma^r_{ki}u^k=-\frac{1}{\rho}g^{rr}\frac{\partial p}{\partial r},
\end{eqnarray}
and then neglecting terms proportional to $\cos^2\theta$ we get
\begin{equation}
u^t\frac{\partial u^r}{\partial t}+u^r\frac{\partial u^r}{\partial r}+\Gamma^r_{tt}u^tu^t+2\Gamma^r_{t\phi}u^tu^\phi+\Gamma^r_{rr}u^ru^r+\Gamma^r_{\phi\phi}u^\phi u^\phi=-\frac{1}{\rho}g^{rr}\frac{\partial p}{\partial r}.
\end{equation}
This equation can be reformulated as
\begin{equation}\label{App_eq_dVdt}
\frac{\partial V}{\partial t}=\frac{\sqrt{1-V^2}\Delta}{\gamma A^{1/2}}\left[-\frac{V}{1-V^2}\frac{\partial V}{\partial r}+\frac{{\cal A}}{r}-\frac{1-V^2}{\rho}\frac{\partial p}{\partial r}\right],
\end{equation}
where $\cal A$, $\Omega$, $\Omega_{\rm K}^+$, $\Omega_{\rm K}^-$, $\tilde{\Omega}$ and $\tilde{R}$ are defined as in \citet{Abramowicz96}
\begin{mathletters}
\begin{eqnarray}
\label{App_def_calA}{\cal A}&\equiv&-\frac{MA}{r^3\Delta\Omega_{\rm K}^+\Omega_{\rm K}^-}\frac{(\Omega-\Omega_{\rm K}^+)(\Omega-\Omega_{\rm K}^-)}{1-\tilde{\Omega}^2\tilde{R}^2},\\
\label{App_def_Omega} \Omega&\equiv&\frac{u^\phi}{u^t}=\frac{2Mar}{A}+\frac{r^3\Delta^{1/2}{\cal L}}{\gamma A^{3/2}},\\
\Omega_{\rm K}^\pm&\equiv&\pm\frac{M^{1/2}}{r^{3/2}\pm aM^{1/2}},\\
\tilde{\Omega}&\equiv&\Omega-\frac{2Mar}{A},\\
\tilde{R}&\equiv&\frac{A}{r^2\Delta^{1/2}}.
\end{eqnarray}
\end{mathletters}
We have neglected the radial components of the viscous stress tensor ($s^{ir}$) as they are negligible for vertically integrated models of accretion disks.

\subsection{Angular momentum conservation}
The angular momentum conservation can be written as
\begin{equation}
\nabla_i(T^i_{k}\xi^k)=0,
\end{equation}
where $\xi^k$($\equiv\delta^k_{(\phi)}$) is the azimuthal Killing vector. After vertical integration we get
\begin{equation}
\Sigma\left(u^t\frac{\partial u_\phi}{\partial t}+u^r\frac{\partial u_\phi}{\partial r}\right)+\nabla_i(S^i_\phi)=0,
\end{equation}
where $S^i_\phi$ is the vertically integrated $(i,\phi)$ component of the viscous stress tensor. We follow \citet{Lasota94} to assume that the only nonvanishing component of $S^i_{\phi}$ is
\begin{equation}
S^r_\phi=-\nu\Sigma\frac{A^{3/2}\Delta^{1/2}\gamma^3}{r^5}\frac{\partial\Omega}{\partial r},
\end{equation}
where $\nu$ is the kinetic viscosity coefficient and $\Omega$ is the angular velocity with respect to the stationary observer [see definition (\ref{App_def_Omega})]. Finally, introducing $\cal{L}$ and $V$ with the equations (\ref{u^t}), (\ref{u^r}) and (\ref{u_phi}) we get
\begin{equation}\label{App_eq_dLdt}
\frac{\partial\cal L}{\partial t}=-\frac{V\Delta}{\gamma\sqrt{1-V^2}A^{1/2}}\frac{\partial\cal L}{\partial r}+
\frac{\Delta^{1/2}}{\gamma\Sigma A^{1/2}}\frac{\partial}{\partial r}\left(\frac{\nu\Sigma A^{3/2}\Delta^{1/2}\gamma^3}{r^4}
\frac{\partial\Omega}{\partial r}\right).
\end{equation}

\subsection{Half thickness evolution}\label{App_section_HalfOpen}
Let us consider the vertical acceleration of the disk surface in the Zero Angular Momentum Observer (ZAMO) system of coordinates.
The disk surface is defined in cylindrical coordinates as
\begin{equation}
z=H(r,t)
\end{equation}
Therefore, the infinitesimal shift in the vertical direction can be expressed in the following way
\begin{equation}
{\rm dz}=\frac{\partial H}{\partial r}{\rm dr}+\frac{\partial H}{\partial t}{\rm dt}.
\end{equation}
Dividing both sides by $\rm dt$ we get
\begin{equation}
\frac{\rm dz}{\rm dt}=-V^{(\theta)}=\frac{\partial H}{\partial r}V^{(r)}+\frac{\partial H}{\partial t}.
\end{equation}
Introducing $U$ and $V$ with the equations (\ref{V^r}) and (\ref{V^theta}) we obtain
\begin{equation}\label{App_eq_dHdt}
\frac{\partial H}{\partial t}=-U\cos\Theta_H-\frac{1}{\gamma}\frac{V}{\sqrt{1-V^2}}\frac{\partial H}{\partial r},
\end{equation}
where $\Theta_H$ is the colatitude angle corresponding to the disk surface ($\theta=\Theta_H$ and $H=r\cos\Theta_H$) at a certain radius.

\subsection{Surface vertical motion}
We follow \citet{Abramowicz97} to assume the form of vertically integrated pressure as
\begin{equation}
P(r,\theta,t)=P_0(r,t)\left[1-\frac{\cos^2\theta}{\cos^2\Theta_H}\right].
\label{Prthetat}
\end{equation}
Let us consider the $\theta$-component of the divergence of stress-energy tensor
\begin{equation}
\nabla_k T^k_{\theta}=0.
\end{equation}
Expanding and vertically integrating we get
\begin{equation}
\frac{1}{\Sigma}\frac{\partial P}{\partial\theta}=-u^k u_{\theta,k}+\Gamma^i_{\theta k}u_iu^k.
\end{equation}
From equation (\ref{Prthetat}) it follows that
\begin{equation}
\left.\frac{\partial P}{\partial \theta}\right|_{\theta=\Theta_H}=\frac{2P}{\cos\Theta_H}.
\end{equation}
Taking both relations into account we have
\begin{equation}\label{App_eq_dUdt}
\frac{\partial U}{\partial t}=\frac{\Delta^{1/2}}{\gamma^2 A^{1/2}\cos\Theta_H}{\cal R}-
\frac{U}{\gamma^2}\left(\frac{V}{(1-V^2)^2}\frac{\partial V}{\partial t}+
\frac{{\cal L}r^2}{A}\frac{\partial\cal L}{\partial t}\right)-
\frac{U}{H}\frac{\partial H}{\partial t},
\end{equation}
where
\begin{mathletters}
\begin{eqnarray}
&&{\cal R}\equiv-\frac{2P}{\Sigma\cos\Theta_H}+
\left({\cal L}^2-a^2(u_tu_t-1)\right)\frac{\cos\Theta_H}{r^2}-
u^r\frac{\partial u_\theta}{\partial r},\label{App_def_calR}\\
&&u_t u_t = \left[\gamma\frac{r\Delta^{1/2}}{A^{1/2}}+\omega{\cal L}\right]^2,\\
&&u^r\frac{\partial u_\theta}{\partial r} = \frac{V\Delta^{1/2}}{r\sqrt{1-V^2}}\frac{\partial}{\partial r}
\left(\gamma Ur\cos\Theta_H\right).
\end{eqnarray}
\end{mathletters}

\subsection{The energy conservation}
From the general form of the energy conservation equation
\begin{equation}
\nabla _i(T^{ik}\eta_k)=0,
\end{equation}
we obtain, in the non-relativistic fluid approximation,
\begin{equation}
\rho\left[u^t\frac{\partial\epsilon}{\partial t}+u^r\frac{\partial\epsilon}{\partial r}\right]+\nabla_i(\eta\sigma^{it})+\nabla_i(u^tq^i+u^iq^t)=0.
\end{equation}
Similarily like in \citet{LL59} (their Eq. 49.4), after vertical integration, it can be reformulated as
\begin{equation}
\Sigma T\left[u^t\frac{\partial S}{\partial t}+u^r\frac{\partial S}{\partial r}\right]=F^+-F^-.
\end{equation}
Where $S$ is the entropy per unit mass, $F^+$ is the local viscous heat generation rate and $F^-$ is the radiative cooling rate (see \citet{Abramowicz96})
\begin{equation}\label{App_def_F^+}
F^+=\nu\gamma^4\Sigma\frac{A^2}{r^6}\left(\frac{\partial\Omega}{\partial r}\right)^2,
\end{equation}
\begin{equation}\label{Aeq:Fm}
%\footnote{Where, equation (\ref{Aeq:Fm}) is the optically thick limit of equation (\ref{F^-}). We indeed use equation (\ref{F^-}) in the whole paper, but equation (\ref{Aeq:Fm}) is retained here for clarity of the derivation.}
F^-=\frac{8\sigma T^4}{3\tau_{\rm{R}}/2+\sqrt{3}+1/\tau_{\rm{P}}}.
\end{equation}
Termodynamical relations give (for $p=p_{\rm rad}+p_{\rm gas}$)
\begin{equation}
T\frac{\partial S}{\partial r}=c_VT\left(\frac{1}{T}\frac{\partial T}{\partial r}-(\Gamma_3-1)\frac{1}{\rho}\frac{\partial\rho}{\partial r}\right),
\end{equation}
\begin{equation}
T\frac{\partial S}{\partial t}=c_VT\left(\frac{1}{T}\frac{\partial T}{\partial t}-(\Gamma_3-1)\frac{1}{\rho}\frac{\partial\rho}{\partial t}\right).
\end{equation}
The appropriate sum of the derivatives of $\rho$ can be substituted using the continuity equation
\begin{equation}
u^t\frac1\rho\frac{\partial\rho}{\partial t}+u^r\frac1\rho\frac{\partial\rho}{\partial r}\approx-\frac{\partial u^t}{\partial t}-\frac{1}{r^2}\frac\partial{\partial r}(r^2u^r)+\frac{U}{r}.
\end{equation}
Taking all together we get
\begin{equation}
c_V\left[\Sigma\left(u^t\frac{\partial T}{\partial t}+u^r\frac{\partial T}{\partial r}\right)-(\Gamma_3-1)T\Sigma\left(-\frac{\partial u^t}{\partial t}-\frac{1}{r^2}\frac\partial{\partial r}(r^2u^r)\right)\right]=F^+-F^-,
\end{equation}
where
\begin{mathletters}
\begin{eqnarray}
&&c_{\rm V}=\frac{4-3\beta}{\Gamma_3-1}\frac{P}{\Sigma T},\label{App_def_cV}\\
&&\Gamma_3-1=\frac{(4-3\beta)(\gamma_{\rm gas}-1)}{12(1-\beta)(\gamma_{\rm gas}-1)+\beta},\label{App_def_Gamma3}\\
&&\beta=\frac{p_{\rm gas}}{p}.
\end{eqnarray}
\end{mathletters}
Ultimately we get
\begin{equation}\label{App_eq_dTdt}
\frac{\partial T}{\partial t}=\frac{1}{\Sigma}\frac{r\Delta^{1/2}}{\gamma A^{1/2}}\left[
\frac{F^+-F^-}{c_{\rm V}}+(\Gamma_3-1)T\Sigma\left(-\frac{\partial u^t}{\partial t}-\frac{1}{r^2}\frac\partial{\partial r}(r^2u^r)\right)\right]-
\frac{V\Delta}{\gamma\sqrt{1-V^2}A^{1/2}}\frac{\partial T}{\partial r}.
\end{equation}

%-------------------------------------------------------------------------------------------------

\clearpage

\begin{figure}
\epsscale{1}
\plotone{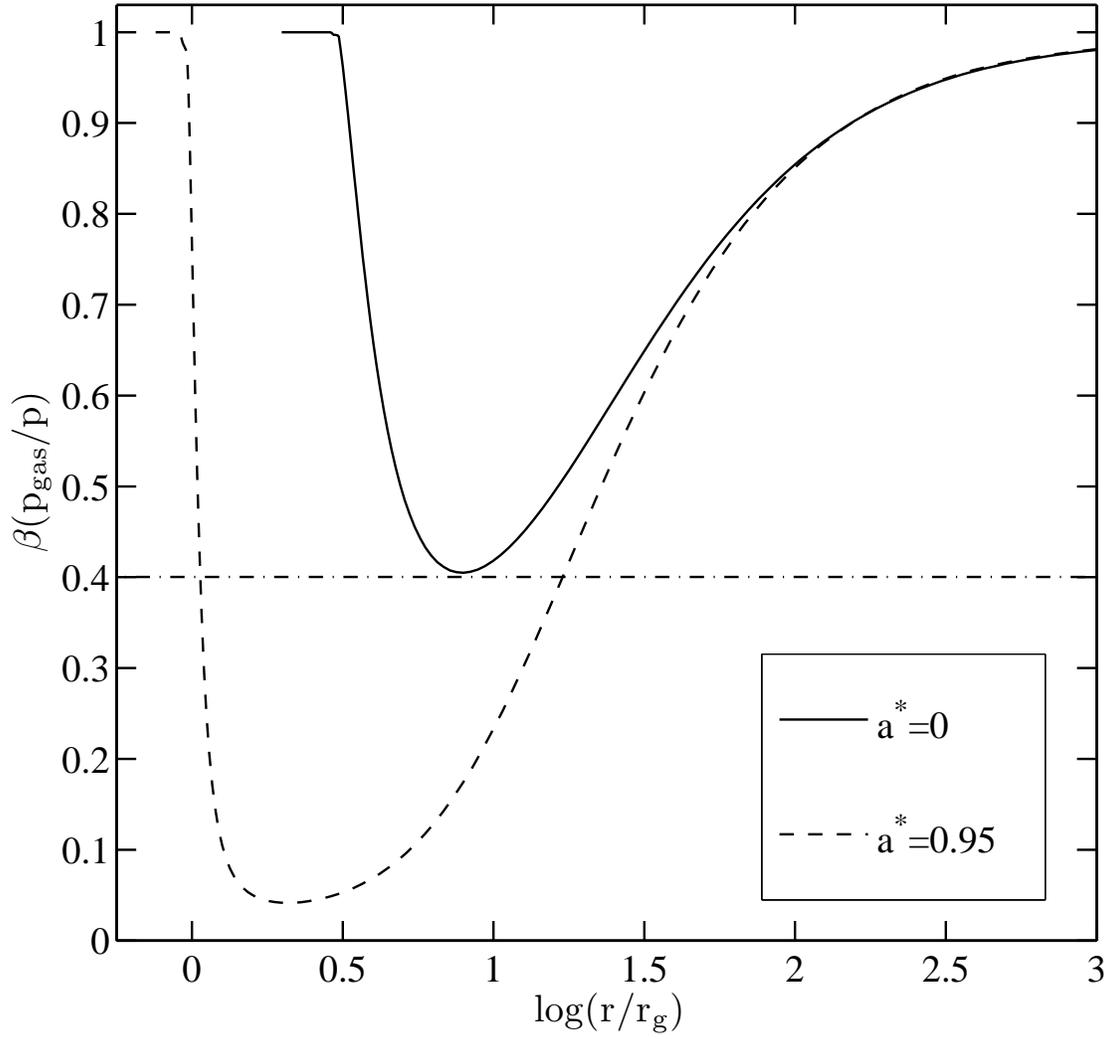}
\caption{Profiles of ratio of gas pressure to total pressure. According to the local analysis of stability \citep[e.g.,][p.155]{Kato_book}, the disk region with $\beta<0.4$ is thermally unstable. \label{fig:beta}}
\end{figure}

\begin{figure}
\epsscale{1}
\plotone{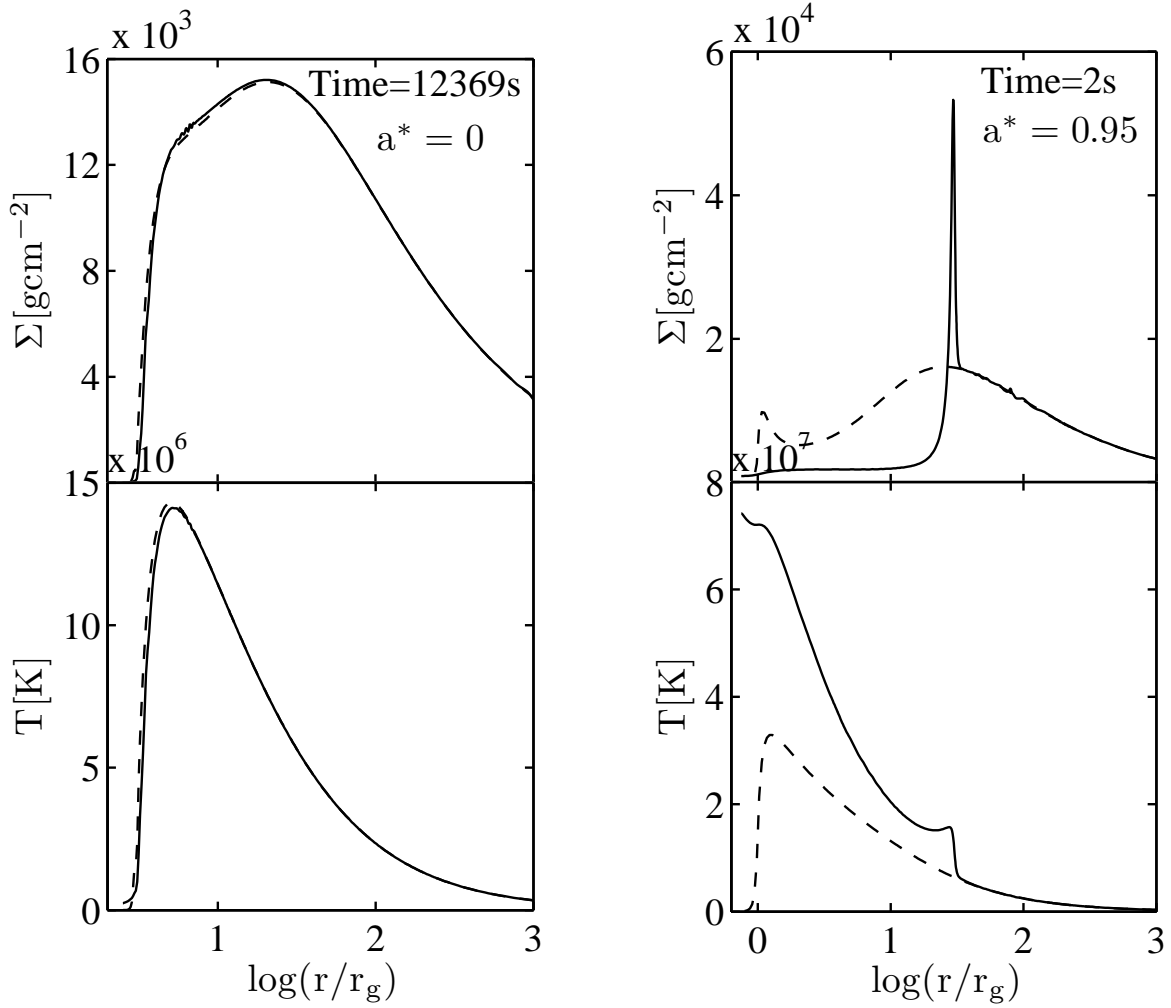}
\caption{Profiles of surface density and temperature of two special cases. Dash lines correspond to the initial state, but solid lines are the relevant snapshots at the certain time. \label{fig:DensTemp}}
\end{figure}

\begin{figure}
\epsscale{1}
\plotone{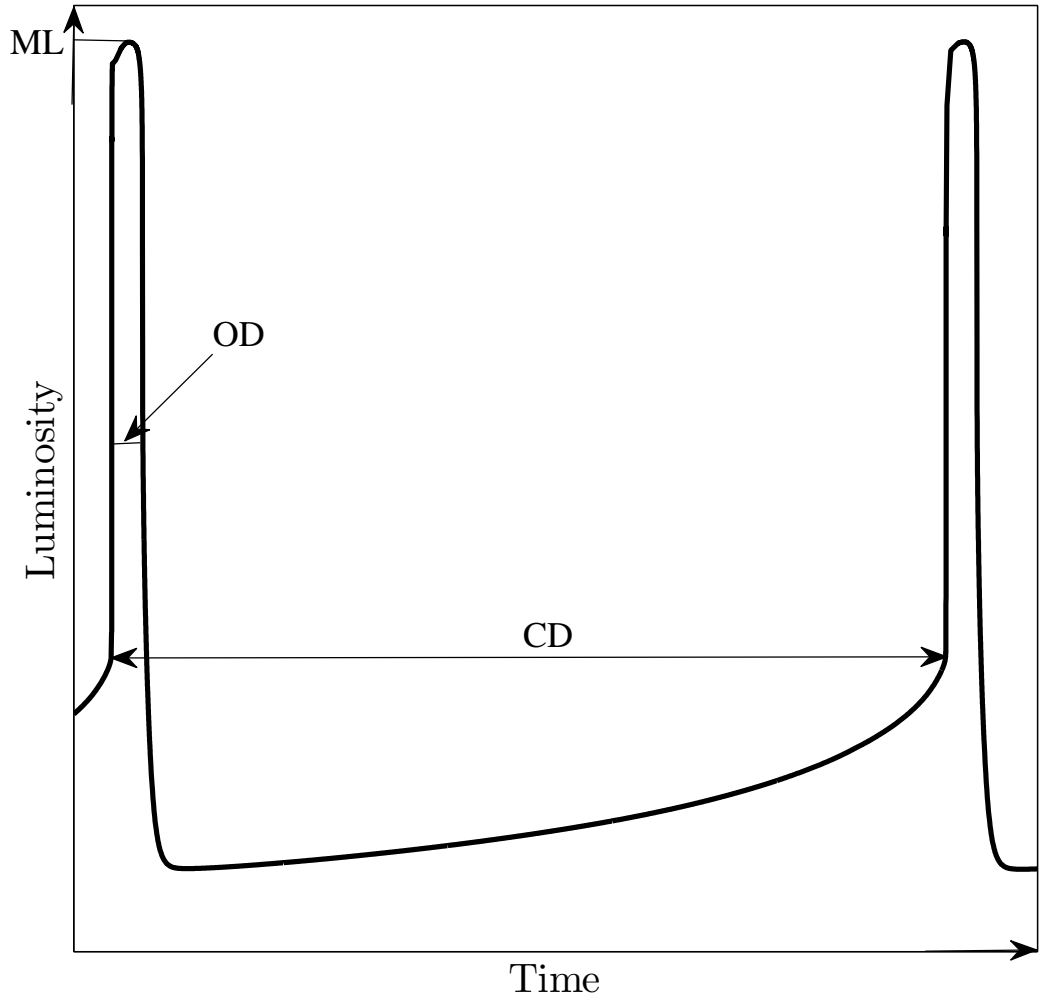}
\caption{Sketch for explaining the definitions of characteristic quantities. \label{fig:sketch}}
\end{figure}

\begin{figure}
\epsscale{1}
\plotone{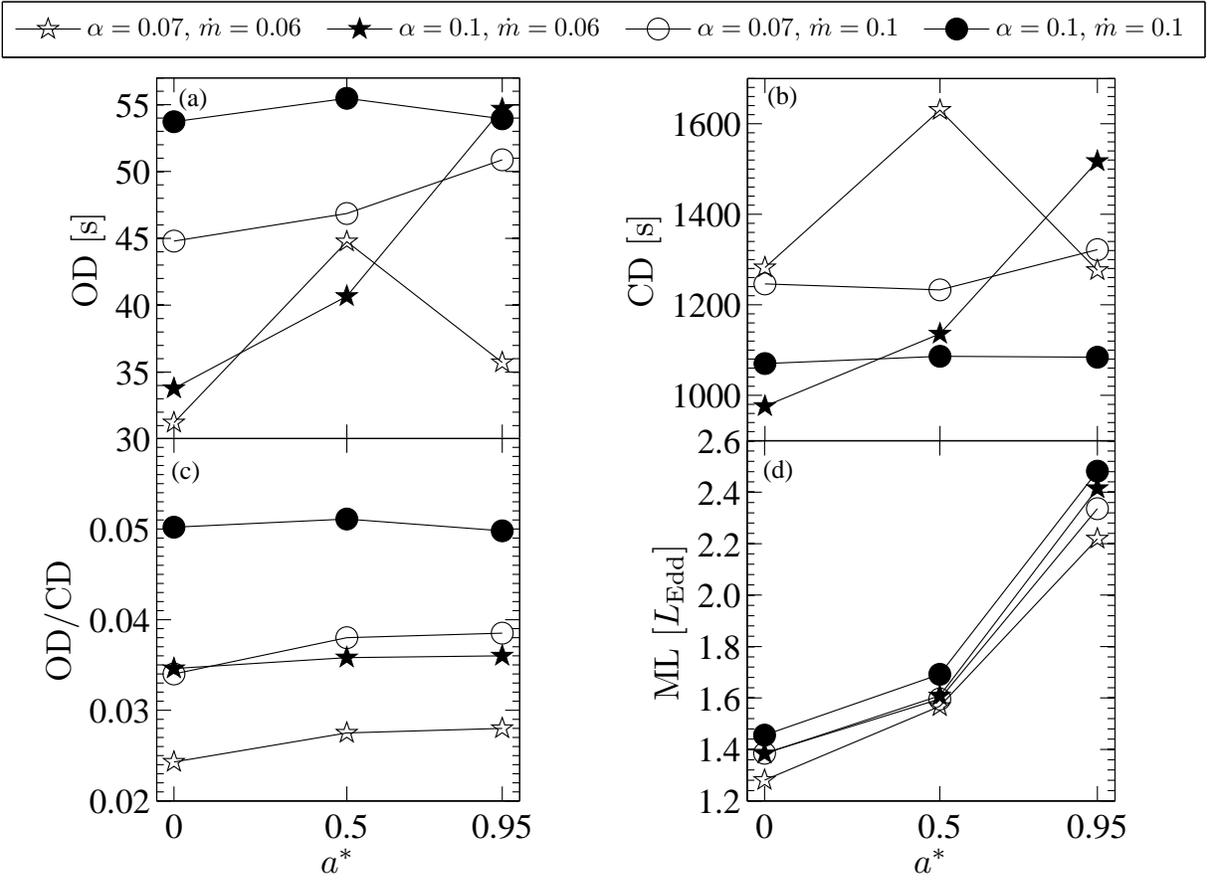}
\caption{Correlation between the characteristic quantities and $a^*$. \label{fig:4plots}}
\end{figure}

\begin{figure}
\epsscale{1}
\plotone{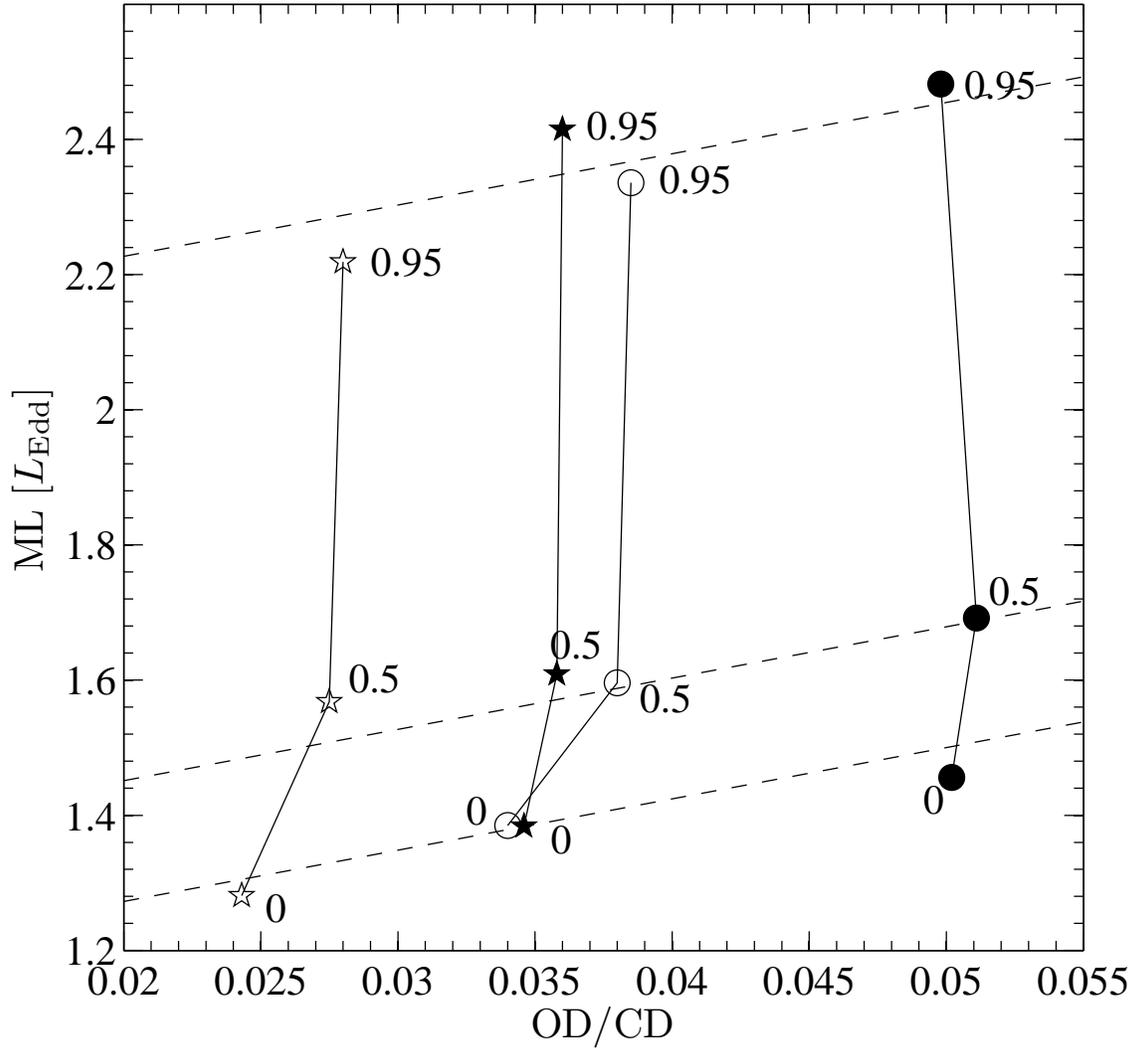}
\caption{Plot of maximal luminosity versus ratio of OD to CD. The numbers near each data point are the values of $a^*$. The meanings of markers are the same as those in Figure \ref{fig:4plots}. The dashed lines are the OD/CD-ML fitted lines for different $a^*$ cases. \label{fig:ML_ODC}}
\end{figure}

\begin{figure}
\epsscale{1}
\plotone{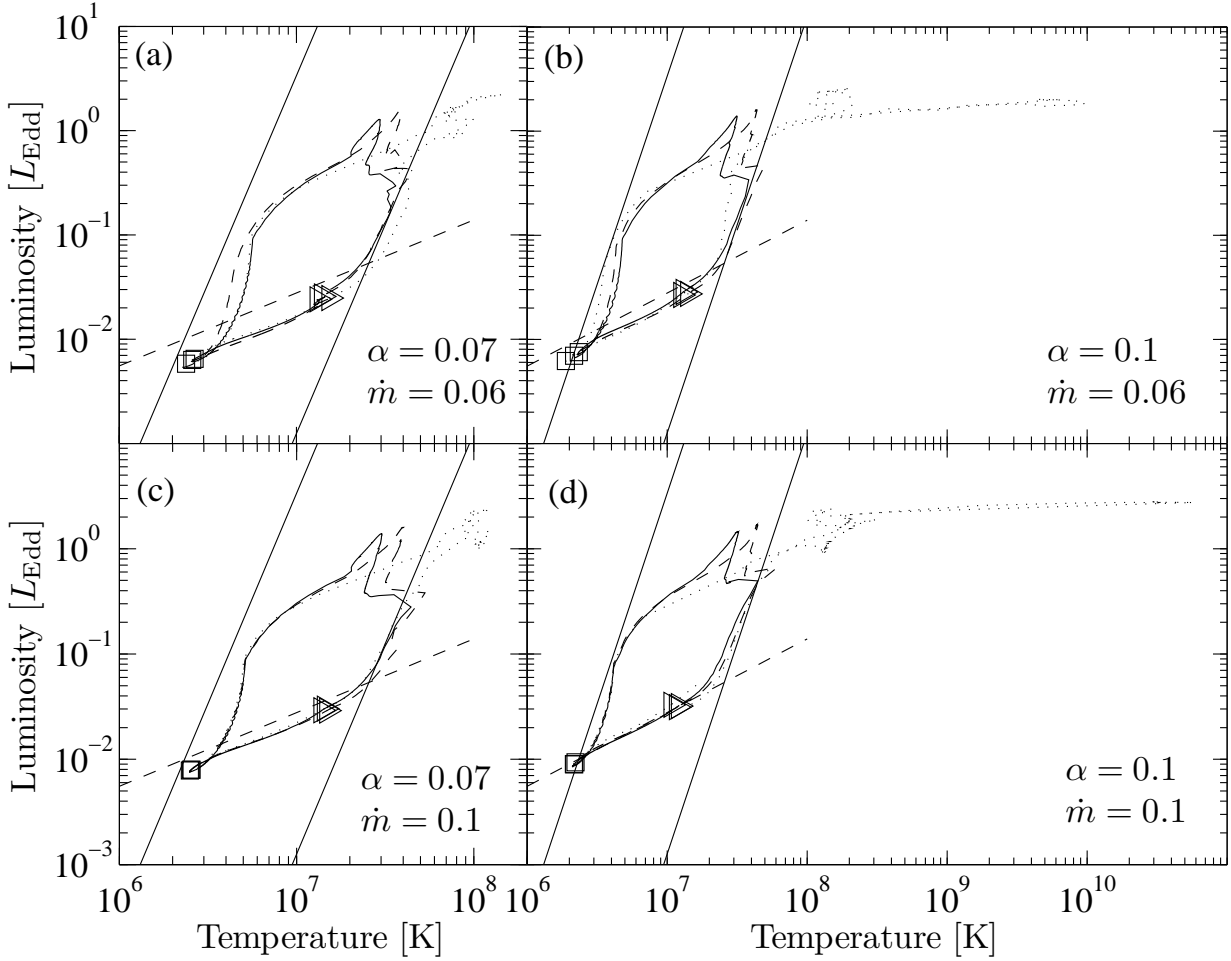}
\caption{Evolution of the last cycle for the groups with different $\alpha$ and $\dot m$ in the maximal disk temperature - luminosity ($T$-$L$) plane. All cycles begin at the triangle points, and then evolve along the curve in the anti-clockwise direction. The square points are used to denote the quarter of cycle durations. The solid straight lines in each panel represent $L_{\rm disk}\propto T^4_{\rm max}$ for a non-spinning BH \citep[see eq. (3) of][]{Gierlinski04}. They are calculated with the color temperature correction $f_{\rm col}=1$ for the left lines and $f_{\rm col}=7.2$ for the right ones. The dashed straight lines denote the $L_{\rm disk}\propto T^{0.7}_{\rm max}$ relation. In each panel, the solid, dashed, and dotted curves are for the different BH spins with $a^*=0, 0.5, 0.95$, respectively. \label{fig:L_MT}}
\end{figure}

\clearpage

\begin{deluxetable}{ccccccccc}
\tabletypesize{\scriptsize}
%\rotate
\tablecaption{Four limit-cycle characteristic quantities of the twelve cases\tablenotemark{1} \label{tab1}}
\tablewidth{0pt}
\tablehead{
\colhead{$a^*$} & \colhead{$\alpha$} & \colhead{$\dot{m}$} & \colhead{OD [s]} &\colhead{CD [s]} &\colhead{OD/CD} & \colhead{ML [$L_{\rm{Edd}}$]}
}
\startdata
0 & 0.07 & 0.06 &31.21$\pm$0.12 (0.38$\%$)\tablenotemark{2}& 1282$\pm$3 (0.23$\%$) &(2.43$\pm$0.02)$\times10^{-2}$ (0.61$\%$)& 1.281$\pm$0.001 (0.07$\%$)\\
0.5 & 0.07 & 0.06 &44.76$\pm$0.32 (0.71$\%$)& 1630$\pm$4 (0.25$\%$) &(2.75$\pm$0.03)$\times10^{-2}$ (0.96$\%$) & 1.568$\pm$0.003 (0.17$\%$)\\
0.95 & 0.07 & 0.06 &35.72$\pm$0.14 (0.38$\%$)& 1276$\pm$10 (0.79$\%$) &(2.80$\pm$0.04)$\times10^{-2}$ (1.17$\%$) & 2.219$\pm$0.018 (0.79$\%$)\\
0 & 0.1 & 0.06 &33.78$\pm$0.12 (0.35$\%$)& 976$\pm$3 (0.31$\%$) &(3.46$\pm$0.03)$\times10^{-2}$ (0.66$\%$) & 1.384$\pm$0.003 (0.18$\%$) \\
0.5 & 0.1 & 0.06 &40.67$\pm$0.21 (0.50$\%$)& 1136$\pm$5 (0.44$\%$) &(3.58$\pm$0.04)$\times10^{-2}$ (0.94$\%$) & 1.609$\pm$0.003 (0.18$\%$) \\
0.95 & 0.1 & 0.06 &54.68$\pm$0.45 (0.81$\%$)& 1517$\pm$10 (0.66$\%$) &(3.60$\pm$0.06)$\times10^{-2}$ (1.47$\%$) & 2.415$\pm$0.008 (0.32$\%$)\\
0 & 0.07 & 0.1 &44.80$\pm$0.15 (0.33$\%$)& 1246$\pm$4 (0.33$\%$) &(3.40$\pm$0.03)$\times10^{-2}$ (0.66$\%$) & 1.385$\pm$0.002 (0.14$\%$)\\
0.5 & 0.07 & 0.1 &46.86$\pm$0.12 (0.24$\%$)& 1233$\pm$3 (0.25$\%$) &(3.80$\pm$0.02)$\times10^{-2}$ (0.49$\%$) & 1.596$\pm$0.003 (0.17$\%$)\\
0.95 & 0.07 & 0.1 &50.87$\pm$0.27 (0.52$\%$) & 1322$\pm$3 (0.23$\%$) &(3.85$\pm$0.03)$\times10^{-2}$ (0.75$\%$) & 2.336$\pm$0.006 (0.23$\%$)\\
0 & 0.1 & 0.1 &53.73$\pm$0.53 (0.98$\%$)& 1070$\pm$4 (0.38$\%$) &(5.02$\pm$0.07)$\times10^{-2}$ (1.36$\%$) & 1.456$\pm$0.002 (0.08$\%$)\\
0.5 & 0.1 & 0.1 &55.47$\pm$0.70 (1.25$\%$)& 1086$\pm$9 (0.83$\%$) &(5.11$\pm$0.11)$\times10^{-2}$ (2.08$\%$) &  1.692$\pm$0.003 (0.17$\%$)\\
0.95 & 0.1 & 0.1 &53.94$\pm$0.47 (0.87$\%$)& 1084$\pm$5 (0.47$\%$) &(4.98$\pm$0.07)$\times10^{-2}$ (1.34$\%$) & 2.482$\pm$0.004 (0.16$\%$)\\
\enddata
\tablenotetext{1}{See \S\ref{Sec_NumExplo} for the definitions of OD, CD, and ML.}
\tablenotetext{2}{The values showed in the columns, OD, CD, OD/CD and ML are collected as "mean value"$\pm$"standard deviation" (relative deviation). }

\end{deluxetable}

\end{document}